Centre de Physique Théorique* - CNRS - Luminy, Case 907
F-13288 Marseille Cedex 9 - France

# Data analysis techniques for resolving nonlinear processes in plasmas : a review

T. Dudok de Wit[1]

Abstract

The growing need for a better understanding of nonlinear processes in plasma physics has in the last decades stimulated the development of new and more advanced data analysis techniques. Progress in this domain has greatly benefited from developments in fluid dynamics and dynamic system theory. This review lists some of the basic properties one may wish to infer from a data set and then presents appropriate analysis techniques with some recent applications. The emphasis is put on the investigation of nonlinear wave phenomena and turbulence in space plasmas.

comp-gas/9611002    22 Nov 1996

*    Unité Propre de Recherche 7061
[1]    also at the Université de Provence;  E-mail : ddwit@cpt.univ-mrs.fr

# 1. Introduction

Plasmas offer a paradigm for nonlinear processes, which in recent decades have received a growing interest. It is now well understood that such processes cannot be properly analyzed with standard linear techniques: Fourier decompositions, for example, often fail to reveal all the relevant information that is contained in a data set. This situation has prompted the development of more advanced analysis techniques that are often based on new concepts. These techniques have greatly improved our understanding of basic physical processes such as wave-wave interactions, intermittency and self-organization.

There are many ways in which a plasma may be considered to behave in a nonlinear manner. Likewise, there exists a large collection of analysis techniques that can be combined in various ways. The aim of this review is to present some of the existing and emerging techniques. Its organization is schematized in Figure 1. It starts by listing some of the basic nonlinear properties one may wish to infer from a data set, and then proceeds with a presentation of appropriate techniques for extracting the information. The emphasis is deliberately put on the investigation of nonlinear wave phenomena and turbulence in space plasmas. The problems encountered in space plasmas, however, are often similar to those met in laboratory plasmas, fluids or dynamical systems. To a large extent, we rely on a blend of physical observation and mathematical techniques developed in communities that have built up intuition over many decades. For that reason, much attention will be given to ideas that issue from other fields, and that may be potentially useful to the space plasma physics community.

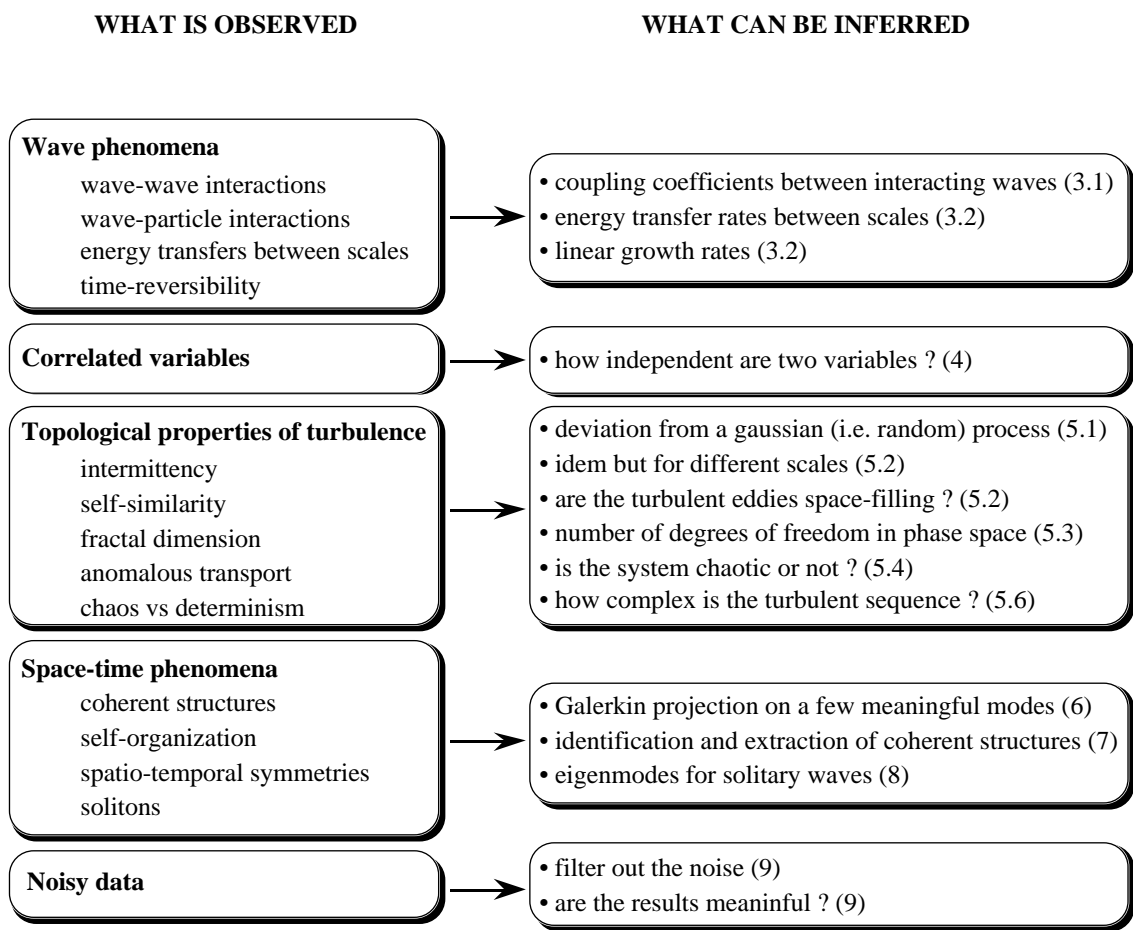

**Figure 1** *Organization of the review with the sections in which the different analysis techniques are presented. Topics related to the estimation of linear properties, such as wavenumber and frequency spectra, are not addressed.*

## 2. Choosing the right technique

Data analysis aims at characterizing physical processes by means of invariants. Most of the work done in this area is heavily influenced by linear descriptions and by decades of Fourier representations. Fourier modes have indeed the advantage to be natural eigensolutions of partial differential equations. This is not true anymore with nonlinear systems, for which there is no such single tool. It is appropriate to say that nonlinear system analysis definitely departs from the more familiar spectral analysis. This situation has prompted the development of variety of new methods.

The first and yet often overlooked step in the choice of an analysis technique is the choice of the right invariants. This requires physical insight: good parameters for hydrodynamic turbulence, such as enstrophy, may not be adequate for 3D magnetohydrodynamic (MHD) turbulence, for which energy and magnetic helicity [*Matthaeus and Goldstein, 1982*] are preferable. The properties of the linearized system usually provide a good starting point. Indeed, one can often assume a weak departure from linearity and accordingly obtain valuable information simply by expanding the solutions of the nonlinear system in terms of its linear eigenmodes. Several powerful techniques are based on this assumption. This approach obviously breaks down for very nonlinear processes, such as strong plasma turbulence. In many cases, however, new empirical invariants are found, such as the energy transfer rate between scales in Kolmogorov's model for turbulence.

Once a set of invariants has been found, it is necessary to determine the ones that can be estimated with reasonable accuracy. Unfortunately, many estimation procedures involve inverse problems, which are traditionally ill-conditioned. This means that different approaches may give different results. In this context, cross-validation between different techniques becomes necessary and the large number of existing tools turns out to be a real advantage. Mastery over no single set of tools is required to address the multiple facets of plasma (and hydrodynamic) turbulence.

Although turbulence remains one of the greatest challenges in physics, progress in modelling and understanding it has been agonizing slow. Needless to say, almost all approaches start from a probabilistic description. Spectral representations are still at the heart of many techniques and benefit from a over fifty years of experience. The theory of dynamical systems has brought new ideas and new tools, but the quantitative connection between dynamical chaos and turbulence is not clear yet. Thus, a recurrent problem in the analysis of nonlinear phenomena is the choice between traditional, well-understood techniques, and novel and more elaborate concepts that one cannot yet interpret with as much confidence.

At the outset it must be said that the analysis of nonlinear phenomena has already been the object of several reviews. Few of these, however, provide an adequate balance between the description of physical phenomena and the application of data analysis tools. The state of the art in the analysis of laboratory plasma turbulence was presented some time ago by *Ritz et al.* [*1988*] but this excellent review predates most of the recent developments in dynamic systems analysis. This gap has partly been filled by two proceedings that are more exclusively devoted to space plasmas [*Lefeuvre et al., 1993; Glassmeier et al., 1995*]. The analysis of chaotic systems (better known as "nonlinear time series analysis") has in comparison received much more attention, albeit rarely in the context of plasma turbulence. Some useful reviews or compilations are those by *Tong* [*1990*], *Drazin and King* [*1992*], *Abarbanel et al.* [*1993, 1995*] and *Ott et al.* [*1994*].

## 3. Nonlinear interactions between waves

The traditional starting point for a nonlinearity analysis consists in building a linear model of the data and then assume a weak departure from linearity. This means decomposing the signal into eigensolutions of the linear system, which are Fourier modes. Besides obvious historical reasons, this approach has been widely advocated because many nonlinear phenomena turn out to be conveniently described in terms of interacting waves (e.g. *Porkolab and Chang* [*1978*]). The decay and the modulational instabilities, for example, which play an important role in space plasmas, involve an interaction between respectively three and four waves. The framework of nonlinear wave interactions is not only valid for monochromatic waves, but can in principle also be extended to broadband processes, such as plasma turbulence.

The basic assumption which underlies linear spectral analysis is that any stationary fluctuating physical quantity may be regarded as a superposition of statistically independent waves. All the relevant statistical information is then contained in the first and second order moments (i.e., the average and the two-point correlation). If, however, some nonlinear or parametric physical processes exist, then the phases of the Fourier transform cease to be uncorrelated and higher order moments contain additional information. The Fourier spectra of these moments are called polyspectra [*Nikias and Raghuveer, 1987*; *Lefeuvre et al., 1995*]. Two cases are distinguished here, depending on whether single- or multi-point measurements are considered.

## 3.1 Wave interactions from single-point measurements

The most commonly used polyspectra are the bispectrum and the trispectrum, which measure the amount of phase-coherence between respectively three and four interacting waves. Higher order spectra such as the tetraspectrum haven't received much attention so far and the corresponding interactions are not expected to be significant [*Krasnosel'skikh and Lefeuvre, 1993*]. The auto-bispectrum of a spatio-temporal wavefield $y(\mathbf{x}, t)$ is defined as

$$B(\omega_1, \omega_2, \mathbf{k}_1, \mathbf{k}_2) = \langle y(\omega_1, \mathbf{k}_1) \, y(\omega_2, \mathbf{k}_2) \, y^*(\omega_1 + \omega_2, \mathbf{k}_1 + \mathbf{k}_2) \rangle \quad (1)$$

where brackets denote ensemble averaging, * complex conjugation, $\omega$ and $\mathbf{k}$ are respectively the angular frequency and the wavenumber vector. This quantity only vanishes if the phases of the components between brackets are independent, which happens when no interaction takes place. The bispectrum is often normalized to the power spectrum and the resulting quantity is called the bicoherence. The latter is bounded between 0 and 1 and gives a direct measure of the strength of the quadratic coupling. All these quantities are generally expressed as a function of $\omega$ only, since the lack spatially resolved measurements prohibits an analysis of the wavenumber dependence.

Numerous successful applications of the bicoherence to plasmas have been reported since the seminal paper by *Kim and Powers* [*1979*] provided a computational framework for it. Using the bicoherence, nonlinear wave interactions were studied in electrostatic turbulence in laboratory plasmas [*Ritz et al., 1989; van Milligen et al., 1995*], in ionospheric phenomena [*LaBelle and Lund, 1992; Pécseli et al., 1993*], in the interference between radio emitters and the ionosphere [*Lagoutte et al., 1989*], in coalescing Langmuir waves in the solar wind [*Bale et al., 1995*], and in strong turbulence at collisionless shocks [*Dudok de Wit and Krasnosel'skikh, 1995*]. The estimation of the trispectrum is computationally more demanding and has received less attention. *Kravtchenko-Berejnoi et al.* [*1995*] recently applied it to simulations of the Zakharov equations.

Polyspectra have now become classical tools for quantifying nonlinear wave phenomena in weak plasma turbulence. They're also applicable to strong plasma turbulence provided that the finite bandwidth of the waves is taken into account. Polyspectra are particularly useful for detecting nonlinear interactions, mode couplings and time-reversibility [*Elgar, 1987*]. They're are also applicable to wave-particle interactions, but the proper interpretation must then be given to the results. Computational and statistical aspects regarding the estimation of bispectra, and to a lesser extent trispectra, now seem to have reached a level of maturity. There are, however, some pitfalls. Like all high order quantities, polyspectral estimates are noise sensitive and often biased. Therefore, a sound assessment of their statistical significance is crucial for the results to be meaningful. In all cases, high stationarity and long records are required. *Pécseli and Trulsen* [*1993*] in addition showed that even linear models can lead to non vanishing bispectra. Finally, it must be stressed that each polyspectrum can only detect a particular order of interaction: the bispectrum detects quadratic interactions, the trispectrum cubic interactions and so on.

## 3.2 Wave interactions from multipoint measurements

The polyspectra discussed above only deal with one signal at a time but their generalization to multipoint measurements is rather straightforward. Deeper physical insight can be gained by regarding each polyspectrum as a coefficient in the Volterra series expansion of the nonlinear physical process [*Billings, 1980; Im et al., 1996*]. One can model the dynamics of the plasma in a quantitative way and the relevant framework is that of system identification [*Bendat, 1990*]. The plasma is modelled in the frequency domain as a black box which has one or more sources (inputs) and responses (outputs). The aim is to characterize the transfer function which links the input signals $u$ to the output signals $y$, using the parametric model

$$\begin{aligned} y(\omega) &= \Gamma(\omega) u(\omega) \\ &+ \sum_{\substack{\omega_1, \omega_2 \\ \omega_1 + \omega_2 = \omega}} \Lambda(\omega_1, \omega_2) u(\omega_1) u(\omega_2) \\ &+ \ldots \end{aligned} \quad (2)$$

The physical interpretation of the complex matrices $\Gamma(\omega)$ and $\Lambda(\omega_1, \omega_2)$ obviously depends on the in- and out-puts. For the typical case where $u$ and $y$ are two probe signals measured at different spatial locations, the linear term $\Gamma(\omega)$ provides the dispersion and the linear growth rate, while the quadratic term $\Lambda(\omega_1, \omega_2)$ gives the wave coupling coefficients and the energy transfer function. The latter is particularly relevant for turbulent processes, since it quantifies the energy exchange rate between different modes in frequency or wavenumber space; see [*Lii et al., 1982*] for an application to neutral fluid turbulence.

Extensive work on nonlinear transfer functions was done by *Ritz et al.* [*1986; 1988*] for fully developed electrostatic plasma turbulence. The computation of cubic and higher order terms, however, was not attempted because of computational reasons. In a more qualitative approach, *Dudok de Wit and Krasnosel'skikh* [*1995*] detected energy transfers in strong plasma turbulence using the phase of the bispectrum. Nonlinear transfer functions certainly provide one of the most powerful tools for quantifying wave interactions in plasmas, and are particularly relevant for two-point measurements.

### 3.3 Other measures of wave interaction

Polyspectra are widely appreciated for their property to easily fit into theoretical frameworks. In particular, they can be directly linked to the Hamiltonian formalism that has been developed by *Zakharov et al.* [*1985*] for the modelling of weak plasma turbulence, There exist, however, other measures of interaction, whose greater conceptual simplicity is generally compensated by a less obvious interpretation. One idea is to filter the fluctuating signal in different frequency bands, and then look whether the envelopes of the filtered waveforms are linearly correlated. *Duncan and Rusbridge* [*1993*] called this the amplitude correlation technique and observed with it energy transfers between different frequencies in drift wave turbulence. The technique is appealing but the underlying theoretical model is not so easy to establish. In a different context, *Price et al.* [*1994*] proposed an ad hoc nonlinear transfer function model to detect a possible driving of the auroral electrojet index by the solar wind or the planetary magnetic field. Finally, one may also consider working in real space rather than in Fourier space. The equivalents of polyspectra in real space are multiple correlations. Apart from a few exceptions [*Costa et al., 1992*], however, multiple correlations are rather cumbersome to use.

### 3.4 Dealing with short records

A recurrent problem common to all spectral techniques is the unavoidable compromise between simultaneous time and frequency (and/or wavenumber) resolution. The celebrated Fourier transform is not always the best way to estimate the spectral content of a signal. More advanced estimators have been developed (see for example [*Haykin, 1979*]) that can deal with shorter records and higher noise levels. Although their discussion falls beyond the scope of this review, we nevertheless mention the autoregressive (AR) method. The idea is to model a time record by a series of $N$ coefficients $\{a_1, a_2, \ldots a_N\}$

$$y(t_i) = a_1 y(t_{i-1}) + a_2 y(t_{i-2}) + \ldots + a_N y(t_{i-N}) + \varepsilon_i \quad (3)$$

where the residuals $\{\varepsilon_i\}$ have to be minimized according to some prespecified criterion. The power spectrum and other spectral quantities can directly be derived from the coefficients $\{a_i\}$ [*Priestley, 1981*]. Parametric models such as the AR model have been advocated for the computation of polyspectra [*Nikias and Raghuveer, 1987*] and may indeed be appropriate for analyzing either broadband turbulence or closely spaced spectral lines.

Parametric models are not only useful for spectral analysis purposes, but can also advantageously be used to enhance nonlinear features of the data. The idea is to "bleach" the data by fitting a linear model and then study the residuals instead of the data. This approach is appropriate when the hypothesis of nonlinearity can be tested against specific models. It is known, however, that residual testing can lead to false claims for nonlinearity when insufficient knowledge is available about the underlying physics [*Luukkonen et al., 1988*; *Theiler and Eubank, 1993*].

### 3.5 Dealing with transients

Another major shortcoming of Fourier representation is the lack of compatibility between the presupposed conditions of homogeneity and stationarity, and the generally transient and intermittent behaviour of plasma turbulence. Fourier transforms often miss small scale features that can be meaningful for a better understanding of turbulence. This situation has spurred the development of the wavelet transform [*Farge, 1992*; *Meyer and Roques, 1993*] which, for a signal $y(t)$ is defined as

$$y(a, \tau) = \int y(t) \frac{1}{\sqrt{a}} h^*\left(\frac{t-\tau}{a}\right) dt \quad . \quad (4)$$

Here $h(t)$ is the analyzing wavelet and $a$ its scale. A classical wavelet is the Gaussian or Morlet wavelet $h(t) \propto e^{2\pi j t} e^{-t^2/2\sigma^2}$ for which each scale is directly related to an instantaneous frequency $f = 1/a$. The continuous wavelet transform can be regarded as a generalization of the windowed Fourier transform, with improved resolution. Standard quantities such as power spectra and polyspectra can be derived from wavelet transforms, and often come out improved.

A distinction should be made between continuous and discrete wavelet transforms. The former are good for analysis purposes since their shape can be tailored to the spectral properties of interest. Continous wavelet transforms, however, are not invertible, and thus cannot serve to reconstruct the wavefield. Discrete transforms on the contrary produce no redundant information but their scales are much harder to relate to usual spectral properties.

The development of wavelet analyses in space plasma physics is still a long way behind that of fluid turbulence and astrophysics. Wavelets were used so far to study the time evolution at different scales in solar wind turbulence [*Mangeney and Grappin, 1993*] and to correlate transient bursts of electromagnetic activity observed during substorms [*Holter, 1995*]. They were also successfully used as a substitute to Fourier transforms in the estimation of polyspectra [*van Milligen et al., 1995*; *Dudok de Wit et al., 1995*]. Wavelet transforms are certainly among the most promising tools for analyzing turbulence and are particularly relevant for self-similar processes, see §5.2.

## 4. Detecting correlations: information theoretic criteria

Nonlinearity is often perceived as the existence of a coupling that should vanish in the linear approximation. Therefore, many studies involve at some stage the computation of the cross-correlation function [*Hooper, 1971*]. This quantity, however, only tells us whether two variables are *linearly* correlated. Two variables $y_1$ and $y_2$ with zero mean are thereby supposed to be independent if their ensemble averages satisfy

$$\langle y_1 y_2 \rangle = \langle y_1 \rangle \langle y_2 \rangle \ . \tag{5}$$

A more general measure of independence is provided by the mutual information. This quantity, which is based on information theoretic criteria (e.g. [*Zurek, 1990*]), measures the degree of independence between two or more variables, using their joint probability distributions. Two random variables are independent if their probability distribution functions satisfy

$$P_{y_1 y_2}(y_1, y_2) = P_{y_1}(y_1) P_{y_2}(y_2) \ . \tag{6}$$

The mutual information between $y_1$ and $y_2$ is defined as

$$I_{y_1 y_2} = H_{y_1} + H_{y_2} - H_{y_1 y_2} \geq 0 \ . \tag{7}$$

where $H_{y_1}$ and $H_{y_1 y_2}$ are the entropies of respectively the probability and the joint probability distributions.

$$\begin{aligned} H_{y_1} &= -\int P_{y_1}(y_1) \log P_{y_1}(y_1) \, dy_1 \\ H_{y_1 y_2} &= -\int P_{y_1 y_2}(y_1, y_2) \log P_{y_1 y_2}(y_1, y_2) \, dy_1 \, dy_2 \end{aligned} \tag{8}$$

Given a measurement of one signal, the mutual information indicates how many bits on average can be predicted about the other one. This positive quantity goes to zero only when the Eq. 6 is satisfied.

Early applications of mutual information appeared in the analysis of chaotic signals [*Fraser and Swinney, 1986; Fraser; 1989, Palus̆, 1993; Prichard and Theiler, 1995*]. *Ikeda and Matsumoto* [*1989*] applied it to chemical turbulence simulations and showed how information is transferred from one wavenumber to another. Some experiments [*Grésillon et al., 1993; Iranpour and Pécseli, 1995*] confirmed the ability of the mutual information to reveal correlations that are not detected by standard means.

The concept of mutual information provides a powerful and elegant framework for analyzing turbulent processes and as such definitely deserves wider acceptance. In particular, the idea of describing turbulence as a process that continuously generates and destroys information [*Shaw, 1981*] seems promising. A weak point of the technique is its sensitivity to the way the integrals in Equation 8 are discretized. This, however, can be partly overcome by choosing generating partitions with irregular grids [*Fraser, 1986*] or by using other density estimators [*Prichard and Theiler, 1995*].

The mutual information is actually only one element among the many information theoretic criteria that are more generally used in the context of information theory and symbolic dynamics. Some further applications are discussed in §5.6.

## 5. Metric invariants and topological properties of turbulence

Turbulence in fluids, gases and plasmas often departs from a Gaussian process and exhibits geometrical properties such as self-similarity, fractal dimension and intermittency or "patchiness". Although these different concepts may seem unrelated, they are deeply rooted in the nonlinear nature of the underlying physics. In the last decade, they have not only greatly modified our perception of turbulence, but also significantly improved our understanding. Most of the work done on metric invariants has concentrated on dynamical systems [*Grassberger et al., 1991; Abarbanel et al., 1993*] but these concepts are now diffusing into other domains. *Marsch* [*1995*] and *Kurths and Schwarz* [*1995*] discuss some of them in the context of space plasmas. For the sake of clarity, I have artificially separated them into different categories according to the properties one may wish to investigate.

### 5.1 Intermittency and Gaussianity

Probabilistic descriptions are at the heart of the characterization of turbulence. A basic quantity that can give significant clues to the nature of underlying physics is the probability distribution function (PDF) of the wavefield amplitude [*Monin and Yaglom, 1975*]. The PDF of a random velocity field is a Gaussian but numerous experiments have reported non-Gaussian distributions with enhanced tails. Many modern turbulence theories also assume a weak departure from Gaussianity. This enhancement of the tails of the distribution is termed intermittency, a property that is perceived as the tendency of a fluctuating signal to be inhomogeneous in space or in time or in both.

Deviations from Gaussianity are usually quantified in terms of higher order moments of the distribution, such as the skewness (third order moment) or the kurtosis (fourth order moment). Their interpretation is straightforward, but the data need to be stationary. Indeed, as shown by *Sethia and Reddy* [*1995*], non-stationarity can be misleading. Note that intermittency implies the existence of a phase coherence between the Fourier modes of the signal, and as such leads to non-vanishing polyspectra (e.g. [*Kim and Powers, 1979*]).

Intermittency has been addressed in connection with enhanced transport in laboratory plasmas [*Jha et al., 1992*]. It has also been discussed in the context of self-organization [*Biskamp, 1993*] and is occasionally viewed as an indication of the existence of coherent structures. In a series of papers, *Burlaga* [*1991, 1993*] later followed by others [*Marsch and Tu, 1993, 1994; Carbone et al., 1994, 1995, 1996; Ruzmaikin et al., 1995*] have analyzed probability distribution functions of magnetic fluctuations in solar wind turbulence. All their observations give evidence for an intermittent behaviour that is characterized by frequent bursts of magnetic activity. Intermittency has also been characterized by means of wavelets but in the context of neutral fluids only [*Meyer and Roques, 1993*].

## 5.2 Self-similarity and energy cascades

The development of dynamical systems theory has led to a renewed interest in the symmetry properties of turbulence. These symmetries, which are broken in the transition from laminar flows to weak turbulence, are thought to be restored in regimes of fully developed turbulence by their chaotic character. Such regimes frequently exhibit a property called self-similarity (or self-affinity), which means that in a given range of scales, the eddies are in a state of local statistical and dynamical equilibrium. Large turbulent eddies decay into series of smaller affine eddies, which in turn decay into smaller ones until viscous dissipation dominates. An important physical invariant here is the energy transfer rate between different scales, which has been the object of numerous investigations since the early work of Kolmogorov and Obukhov (see for example *Frisch* [*1995*]).

The investigation of self-similar properties is usually based on considerations of PDFs at different scales. The appropriate tool is the structure function [*Monin and Yaglom, 1975*]

$$S_v(p, x) = \langle |v(l) - v(l+x)|^p \rangle \quad , \qquad (9)$$

which quantifies the $p$'th order moment at a characteristic scale $x$. In practice, the Taylor hypothesis of frozen-in turbulence is invoked to replace the spatial dependence, which is difficult to measure, by the temporal dependence.

Likewise, the velocity field can with some assumptions be replaced by the magnetic field. For a self-similar wavefield with constant spectral energy transfer, a universal scaling law holds

$$S_v(p, \tau) \propto \tau^{\alpha(p)} \quad , \qquad (10)$$

where $\tau$ is now the characteristic time scale and the scaling exponent is $\alpha(p) = p/3$. Any deviation from this linear dependence indicates that the wavefield is locally intermittent. By comparing this dependence with theoretical models, one can estimate the degree to which the turbulent eddies fill the 3D physical space [*Paladin and Vulpiani, 1987*]. Self-similar properties of turbulent processes are also often characterized by the slope of the wavenumber spectrum, i.e.,

$$P(k) \propto k^\gamma \quad , \qquad (11)$$

where for a given energy transfer model, the spectral index $\gamma$ is uniquely related to the scaling exponent $\alpha(p)$ of Eq. 10. The structure function sometimes appears in its normalized form

$$A_v(p, \tau) = \frac{S_v(p, \tau)}{S_v(2, \tau)^{p/2}} \qquad (12)$$

which, for Gaussian statistics, should not depend on $\tau$.

Most of the experimental work done in plasma physics has been centered on solar wind turbulence (see *Tu and Marsch* [*1995*] for a review). *Burlaga* [*1991*] first pointed out the striking similarity between neutral fluid and plasma turbulence in the slow solar wind. The spectral index of the latter is often closer to the celebrated -5/3 value found in neutral fluids than to the -3/2 value predicted by MHD theory. This result may have to do with intermittency effects [*Carbone, 1994*]. The statistical properties of the solar wind turbulence are dominated at small scales by large and sparsely distributed fluctuations. Such eddies are not space-filling but occupy a fraction only of the 3D space. Estimates range between 2 and 3, thus suggesting a mixing between 3D eddies and sheet-like structures [*Carbone et al., 1996*]. The scaling properties of the structure functions in the solar wind support the multifractal nature of MHD turbulence. Hierarchical structures have not only be observed in plasmas, but also appear, for example, in microwave bursts caused by solar flares [*Krüger et al., 1994*].

Besides their intrinsic interest, topological invariants play a key role in our understanding of turbulence. Through systematic comparisons between theoretical predictions and observed invariants one can address various questions such as the presence of possible symmetries, energy cascades and eddy interactions. *Marsch and Tu* [*1992*], for example, showed that the fluctuations in the solar wind are not just remnants of coronal events. A comparative study of different regions in space should, in this sense, be enlightening. There are, however, some open questions regarding the interpretation of solar wind data. Indeed, magnetofluids are often highly structured and therefore much less amenable to a comprehensive theoretical description as neutral fluids with the Navier-Stokes equation. Some basic differences arise from the presence of magnetic fields, that break symmetry, and the frequent coexistence of different types of waves. This relative complexity of plasma turbulence constitutes a major obstacle to a more quantitative understanding of its statistical properties. *Dudok de Wit and Krasnosel'skikh* [*1996*], for example, showed that the occasional occurrence of discrete wave packets in the turbulent wavefield can completely alter the self-similar properties deduced from structure functions. Furthermore, it is not clear whether solar wind turbulence is always fully developed or if Taylor's hypothesis of frozen-in turbulence is valid for fluctuations that are not necessarily Alfvénic.

The computation of the structure function is straightforward but its statistical significance is difficult to assess. *Tennekes and Wijngaard* [*1972*] showed that very long records are needed to estimate high order moments. Unfortunately, these high order moments (typically p>8) are needed to discriminate concurrent models for intermittency. Recently, evidence was given that wavelets are better suited for a multifractal description of self-affine phenomena [*Muzy et al., 1993*]. Indeed, wavelets are appropriate for investigating statistical properties at different scales [*Mahrt, 1991*] and in addition provide access to the singularity spectrum [*Kevlahan and Vassilicos, 1994*]. These new concepts, which have emerged in hydrodynamics, may be equally relevant to plasma turbulence.

In spite of the difficulties that arise in the interpretation of self-similar signatures in plasmas, the concept of self-similarity remains simple and can be informative. Many applications have been found in the Earth's sciences and several potential ramifications exist for plasma physics. Two of these are the investigation of scale-invariance in nonlinear wave interactions [*Zakharov, 1984*] and the generalization of the structure function to vectorial quantities. The properties of solenoidal fields, for example, impose constraints on the structure functions [*Monin and Yaglom, 1975*] that could be tested.

Another relevant and recent application is the analysis of self-organized criticality, from which well-defined self-similar properties result. Self-organized criticality arises in many natural systems that are marginally stable [*Bak et al., 1990*] and may be a good candidate for explaining anomalous transport in plasmas [*Newman et al., 1996*]. The large amount of attention devoted to this concept certainly has to do with the fact that it provides one of the few bridges between the topological properties of experimental systems and the physics of low-order deterministic processes.

## 5.3 Low-dimensional systems and fractal properties

It is now widely appreciated that even complex systems such as plasma turbulence can be governed by low-dimensional deterministic models. This property, called deterministic chaos, has prompted an intensive search for low-dimensional phenomena. Indeed, many of the powerful concepts developed in the context of dynamical systems can a priori be applied to turbulence. Some fruitful applications are the onset of chaos in relation with weak turbulence (e.g. Taylor-Couette flows in neutral fluids), and the analysis of chaotic advection. The applicability of dynamical systems theory, however, to fully developed turbulence with many degrees of freedom is less obvious and often still controversial. Many concepts such as attractors and diverging trajectories in phase space (see next section) are difficult to relate to well known physical properties. Some critical discussions on this can be found in [*Crutchfield and Kaneko, 1988; Daglis and Tsinober, 1993; Frisch, 1995*]. Most of these discussions, however, focus on dissipative systems that are relevant for neutral fluids. Plasmas are in comparison more closely related to the theory of Hamiltonian systems and this has deep consequences on their dynamical properties [*Sagdeev et al., 1988*].

Dimension calculations have been quite fashionable for a decade and low-dimensional attractors have been found in many experiments. Numerous articles (*Ott et al.* [*1994*] list the main ones) have been devoted to them, partly because their estimation is challenging and prone to errors. It is now known that even random signals can give rise to finite dimensions. Two of the most widely studied measures of stochasticity are the correlation dimension and the fractal dimension. Both define the dimension in phase space of the attractor towards which all trajectories converge. Their values give some indication of the effective degrees of freedom in phase space. Dimension estimations must be done in two steps. In the first one, the phase space is reconstructed from the time series using different delayed versions of the same time series. In the second step, this so-called embedded map is analyzed. Fundamental limitations of the technique, however, prohibit reliable assessments of dimensions higher than 5.

Early dimension estimates in plasmas were based on fluctuation data from fusion devices (see [*Prado and Fiedler-Ferrari, 1991*] for a review). Most values were larger than 3 or beyond resolution. Work done on magnetospheric [*Roberts, 1991; Pavlos et al., 1992, 1994*] and ionospheric [*Takalo et al., 1994; Wernik and Yeh, 1994*] data also gave evidence for the existence of relatively low-dimensional attractors. All these results show a large scatter, which may originate from a mixing of more or less organized states.

Although there has occasionally been evidence for the existence of strange attractors, the presence of low-dimensional systems often remains questionable. It must be stressed that, in contrast to most laboratory plasmas (in which there has occasionally been evidence for the existence of deterministic chaos [*Skiff et al., 1987*; *Klinger et al., 1993*]), space plasmas are rarely autonomous since they have no boundary conditions. This means that external driving sources may actually strongly influence their phase space properties.

## 5.4 Chaotic behaviour and Lyapunov exponents

In addition to the static measures of stochasticity described above, there also exist dynamic measures. A natural property of chaotic systems is the exponential divergence of originally nearby trajectories. Any volume in phase space will tend to contract along certain directions and expand along others. This rate of change is usually quantified in terms of Lyapunov exponents [*Ott et al., 1994*], whose estimation proceeds from a reconstruction of the phase space and a subsequent analysis of trajectories. The use of Lyapunov exponents is motivated by their connection to the Kolmogorov-Sinai entropy, which characterizes the degree of complexity of trajectories. A system is said to be chaotic if at least one of the exponents is positive.

Lyapunov exponents have become classical tools for studying dynamical systems but computational problems often restrict their extraction to the largest one only. Early applications to low-dimensional experimental systems were devoted to Rayleigh-Bénard convection [*Eckmann et al., 1985*]. *Pavlos et al.* [*1992*] confirmed the existence of at least one positive exponent in ionospheric and magnetospheric data. The requirement for having long and stationary data sets may explain the lack of applications to experimental data. Furthermore, the relevance of Lyapunov exponents to spatially extended systems such as plasma turbulence is debatable.

## 5.5 Anomalous transport and topological properties of wavefields

Theoretical investigations and an increasing number of experiments in space and laboratory plasmas now lend support to the non-Euclidean topology of plasma turbulence. Magnetic field lines, for example, have been found to decompose into fractal sets of regular islands surrounded by stochastic regions. This behaviour has recently received a great deal of attention since it may strongly alter the transport properties of the medium. It is connected to the theory of Lévy flights, which cause non-diffusive (i.e., anomalous) transport [*Solomon et al., 1994*; *Zimbardo et al., 1995*]. Particle percolation across stochastic magnetic field lines, for example, may explain plasma transport from the solar wind into the magnetosheath.

No experimental investigation has succeeded so far in clearly identifying anomalous transport processes in space plasmas. This can be attributed to the numerous difficulties that arise in such an ill-conditioned problem. The recent interest in self-organized criticality has brought new ideas for studying anomalous transport, see §5.2. Another motivation for studying such processes resides in the striking similarity between turbulent eddies and larger objects such as galaxies, to which percolation concepts [*Shandarin and Zeldovich, 1989*] are equally applicable. In a related context, *Milovanov and Zelenyi* [*1993*] investigated the topological properties of braided magnetic field lines in the photosphere and proposed a model to connect the observations to the dynamics of sunspots. The importance of such issues for our understanding on transport will certainly foster new developments in a close future.

## 5.6 Complexity and symbolic dynamics

This title may sound rather abstract, but it hides a series of powerful and yet badly known concepts. In the last decade, deep insight has been gained in the understanding of discrete dynamical systems (such as iterated maps) by considering time sequences as strings of symbols and characterizing the latter with complexity measures [*Wackerbauer et al., 1994*]. Some of these concepts are, with some caution, applicable to real data. The heuristic idea is to encode a measured time series into a sequence of symbols. One of the simplest examples is a binary encoding, in which '1' is assigned to values that exceed the mean and '0' to others. Information theory then provides a variety of complexity measures, which reveal for example the degree of randomness of the sequence or the complexity of the algorithm that was needed to generate it. One may eventually try to identify the grammar that built the sequence but this formidable task is at present still restricted to the simplest systems we know.

It should be stressed these complexity measures are connected to quantities discussed before (in particular the Kolmogorov-Sinai entropy, the Lyapunov exponents and the mutual information), the basic difference lying in a different viewpoint.

Direct applications of symbolic dynamics to plasma data still seem to be lacking. The closest application was devoted to solar flare events [*Schwarz et al., 1993*] and gave evidence for the existence of a nonlinear deterministic process. Complexity measures obviously do not convey the same meaning with real turbulent sequences as with discrete systems for which they were originally developed. What is of interest, however, is not as much their quantitative interpretation as a comparison between the values that issue from different regimes or from different regions.

## 6. Nonlinear effects in space and in time

Most of the techniques discussed so far apply to records that depend on one variable only, generally time or space. They are therefore called univariate. The analysis of records which depend on more than one variable, such as spatio-temporal data, is much more demanding since it formally requires multivariate techniques. There is now a growing interest for such techniques but most of the developments are still restricted to linear processes only [*Russell, 1988; Lefeuvre et al., 1993*]. Nonlinear spatio-temporal processes are in comparison very far from a full understanding. There have traditionally been three ways to go around this.

The simplest solution consists in handling spatio-temporal data as if they were a collection of independent time or space series. This allows standard techniques to be applied but significant information about the space-time dynamics is likely to be lost. The transition from a temporal synchronization of modes to a spatial synchronization, for example, cannot be analyzed that way. At the other extreme, we have a small collection of multivariate nonlinear techniques [*Gifi, 1990*] which are unfortunately of little use to the sort of problems we encounter in plasma physics. A notable exception is the nonlinear transfer function approach, discussed in §3.2.

A compromise between the two previous solutions consists in reducing the spatio-temporal data to a series of separable modes, which are thereafter subjected to standard techniques. What matters here is the choice of the modes. Except for a few special cases, such as travelling solitons, there is no a priori reason for selecting a particular basis. A method that has nevertheless gained wide acceptance is the proper orthogonal decomposition, also known as the Karhunen-Loève expansion or the biorthogonal decomposition. This method is related to the minimum variance analysis and provides a unique decomposition into a set of orthonormal modes

$$y(\mathbf{x},t) = \sum_k A_k\, f_k(\mathbf{x})\, g_k(t) \qquad (13)$$

with

$$\left\langle f_k(\mathbf{x}) f_l^*(\mathbf{x}) \right\rangle = \left\langle g_k(t) g_l^*(t) \right\rangle = \delta_{kl}\,. \qquad (14)$$

The modes $f_k(\mathbf{x})$ and $g_k(t)$ are eigensolutions of the two-point correlation matrices and capture different spatio-temporal features. They were shown to be useful for describing spatio-temporal symmetries and bifurcations in shear flows [*Aubry et al., 1992*]. Symmetries typically arise when there are travelling waves. The latter were recently studied in the transition to turbulence in a plasma device [*Madon and Klinger, 1996*]. It must be pointed out, however, that the proper orthogonal decomposition merely provides a linear projection of the spatio-temporal dynamics, so that concurrent nonlinear effects may not be well separated.

In a similar approach, wavelets were used to extract modes from a turbulent wavefield [*Meneveau, 1991*]. Applications have so far concentrated on analytical and numerical solutions of the Navier-Stokes equation but there is no doubt that wavelets are equally relevant for the analysis of complex spatio-temporal phenomena in plasmas. As will be shown below, both the wavelet transform and the proper orthogonal decomposition are suited for the analysis of coherent structures.

## 7. Coherent structures and self-organization

One of the most fascinating manifestations of nonlinearity is the emergence of large-scale and long-lived structures in high Reynolds number turbulence. The large interest devoted to this phenomenon, called self-organization, has led to the discovery of coherent structures in a variety of experiments. Classical examples are found in neutral fluids [*Lumley*, 1991], but coherent structures have also been observed in plamas as well [*Zweben, 1985; Lu et al., 1993; Horton, 1996*]. Such structures are expected to be effective in mediating transport in plasmas but their role is still largely underestimated. It is interesting to point out the role coherent structures have played in gradually shifting our attention from traditional statistical analysis techniques to more deterministic approaches.

The study of coherent structures provides a fine example in which the lack of adequate analysis techniques has impeded a better understanding of the observations. Early work was essentially based on visualization [*Cantwell*, 1981] and later on cross-correlations between probe signals [*Zweben, 1985*]. Meanwhile, more advanced techniques have emerged but there still remains an important need to develop quantitative and objective tools. Each of the different techniques that follow assumes different definitions of what a coherent structure is, and have their own advantages and restrictions.

### 7.1 Polyspectral analysis

The presence of coherent structures implies a phase-locking between modes that are characterized by different scales. Such phase-lockings could in principle be detected by means of polyspectral techniques (see § 3), which have therefore been proposed as indicators for self-organization [*van Milligen et al., 1995*]. There exist, however, other effects which can cause non vanishing polyspectra. Furthermore, no clear indication on the shape or the lifetime of the structures can be obtained that way.

### 7.2 Conditional averaging

The heuristic idea of conditional averaging (or conditional sampling) is to add ensembles of spatio-temporal data in such a way that random fluctuations are averaged out but not coherent structures. This is done by averaging different ensembles of the wavefield $y(\mathbf{x}, t)$ provided that a certain condition on the amplitude is satisfied

$$y_c = \langle y(\mathbf{x},t) \mid y_o - \varepsilon < y(\mathbf{x}',t') < y_o + \varepsilon \rangle \quad (14)$$

This statistical technique, known for some decades in hydrodynamics, was extensively applied to 2D electrostatic turbulence in plasma devices [*Øynes and Rypdal, 1993*; *Nielsen et al., 1994*]. It is appropriate for visualizing eddies, although quantitative analyses may be biased by spurious effects such as vortex trapping [*Horton et al., 1994*]. A clear advantage of the conditional sampling is its ability to reveal the spatial structure of eddies with two probes only, one which is kept at a fixed position (the one on which the condition is imposed) and one which is displaced. The drawback is that it only provides an averaged shape which may hide fine scales.

### 7.3 Proper orthogonal decomposition

This technique, which separates a multivariate signal into orthonormal modes, was already mentioned in section 6. Recent applications to laboratory plasmas revealed its ability to identify coherent structures [*Benkadda et al., 1994*] and to separate concurrent physical processes such as a superposition of waves [*Dudok de Wit et al., 1994*]. The proper orthogonal decomposition, compared to other decompositions, minimizes the number of modes necessary to represent the data, and thus offers a very compact representation of spatio-temporal signals. Strong points are its objectivity and the possibility to separate the coherent structures from the randomly fluctuating background. To be effective, however, the technique requires good spatial and temporal resolution. Furthermore, the fact that the modes are not known a priori but are data-adaptive makes them occasionally hard to interpret. Some more elaborate techniques have been developed, with other normalizations for the modes (see for example [*Uhl et al., 1995*]).

### 7.4 Wavelet transforms

One of the most successful fields of investigation of wavelet transforms is the analysis of coherent structures in turbulence. Wavelets have been extensively used to identify and extract structures from 2D or 3D fluid turbulence simulation data [*Farge, 1992*; *Berkooz et al., 1992*]. They share the same properties as the proper orthogonal decomposition, but the data are now decomposed into elementary modes whose shape can be controlled. Applications of wavelets to plasma turbulence are unfortunately not as far along. A reason for this may be that the analysis of coherent structures requires spatio-temporal wavelets, whose formalism is at a less advanced stage than that of purely temporal (or spatial) wavelets. Rapid progress, however, is expected to occur in this field.

### 7.5 Neural networks

Latest developments in data analysis also include neural networks, which are known for their ability to detect certain pre-determined patterns or waveforms [*Taylor, 1992*]. Neural networks provide a nonlinear multidimensional mapping which, after being trained on a reference data set, can detect specific patterns such as spectral lines or vortices. A recent study suggested the possibility to detect whistler wave packets in simulation data [*Muret and Omidi, 1995*]. This approach, however, provides little guidance to the physical interpretation of the data.

## 8. Solitons

The spatio-temporal analysis techniques discussed so far do not presuppose any specific molecular entity or coherent structure. A notable exception is the case of solitons or solitary waves. In plasma physics, one often deals with systems that are Hamiltonian (i.e., non dissipative) or close to Hamiltonian. Some of these systems are known to admit soliton-like solutions [*Petviashvili and Pokhotelov, 1992*]. The latter result from a competition between nonlinearity and dispersion, and have been observed in various media, including plasmas. Although solitons can be described in terms of Fourier modes, the latter do not provide any deeper insight because their amplitudes are not conserved. The right solution is to decompose the nonlinear wave motion into a superposition of nonlinearly oscillation modes, whose amplitudes are the invariants of interest. These oscillation modes are obtained from the inverse scattering transform (IST) [*Ablowitz et al., 1974*] which can be interpreted as a nonlinear generalization of the Fourier transform to the solution of nonlinear partial differential equations.

The IST has been applied to different models for weakly nonlinear and weakly dispersive media. Applications to experimental data, however, are just emerging. *Osborne and Petti* [*1994*] successfully described the dynamics of solitons in shallow-water waves using solutions derived from the Korteweg-de Vries equation. They clearly showed how the amplitudes of the oscillation modes are conserved. In a series of papers, *Hada et al.* [*1993*] and *Mjølhus and Hada* [*1995*] advocated the IST for analyzing nonlinear MHD waves in space plasmas.

There is no doubt that the IST is a promising tool for analyzing and describing in a compact way solitary waves in plasmas. A large number of nonlinear wave equations are known to be integrable in terms of it. Two reasons, however, have prevented the IST from gaining wider acceptance. First, a model of the system is needed to decompose the data into oscillation modes. The derivative nonlinear Schrödinger equation is a good candidate for Langmuir solitons and obliquely propagating nonlinear Alfvén waves whereas solitary waves in unmagnetized plasmas are better described by the Korteweg-de Vries equation. Theoretical developments such as a generalization of the analytical solutions to more than one dimension are still in progress. A second obstacle to a broader diffusion of the IST is its computational investment. New numerical techniques, however, are now becoming available [*Osborne, 1993; Candela, 1994*].

## 9. Dealing with noise

An recurrent problem in the analysis of experimental data is the separation between "signal" and "noise". This separation is already known to be a delicate and often subjective task when the processes under consideration are linear. The situation is much worse with nonlinear processes since what may be rejected as noise often contains pertinent information about the nonlinearity. A conservative definition of the term "noise" would involve the fraction of the data that is not explained by the (nonlinear) model. This raises two important questions: 1) how do we know whether an experimental signal is stochastic, deterministic or a combination of the two, and 2) how does noise affect the estimates that have been discussed before, such as the correlation dimension or the polyspectra ?

Although the first question falls beyond the scope of this review, some solutions are worth mentioning since they exploit the nonlinear nature of the underlying physical processes. The state of the art has been presented by *Drazin and King* [*1992*], *Casdagli and Eubank* [*1992*] and *Weigend and Gershenfeld, 1994*]. Two other recent references are [*Wayland et al., 1993; Schreiber and Kaplan, 1996*]. All the techniques share the idea of building a nonlinear model of the data, and thereafter testing the predictions against the data.

The second question has stimulated the development of various tests. A useful one consists in comparing the results to those obtained from surrogate data under the same conditions [*Theiler et al., 1992*]. A surrogate signal is built by computing the Fourier transform, randomizing the phases and subsequently inverting the transform. This procedure preserves first and second order quantities such as the power spectrum, but not higher order ones such as the bispectrum, which vanish. Whenever a surrogate data set yields results that are comparable to the original ones, the inferred nonlinear invariants are very likely to be meaningless. This test has recently been improved by combining it with information theoretic criteria [*Paluš, 1995*]. Such tests are unavoidable when it comes to interpreting with confidence quantities such as Lyapunov exponents or fractal dimensions.

## 10. Conclusions

There is no doubt that the analysis of nonlinear processes in plasmas is still open to considerable progress. Future developments, however, are probably not as much to be expected in the emergence of new concepts as in the suitable application of techniques that have already been used in other fields. Indeed, many techniques have reached a fairly high degree of maturity and sophistication but their application to plasmas is still in its infancy. What is now missing is a stronger cross-disciplinary interaction, which, I believe, is essential for fostering our understanding on such complex phenomena as plasma turbulence. In particular, many of the tools developed for nonlinear and chaotic systems are still waiting for a more straightforward link with the interpretation of the underlying physics : showing evidence for nonlinearity is good but identifying the nonlinear process is much better… For obvious reasons, the most successful tools are those that readily fit into a theoretical framework and thereby bridge the gap between physics and data analysis tools. Two examples are self-organized criticality in relation with self-similarity, and polyspectra in relation with wave interaction models.

The techniques presented here are likely to gain wider acceptance once their advantages over linear descriptions will be recognized. Polyspectra are particularly promising for providing information on nonlinear wave phenomena such as Langmuir wave interactions or nonlinear wave steepenings. Multipoint experiments such as AMPTE should benefit from techniques such as nonlinear transfer functions. The relevance of chaotic data analysis tools for space plasma turbulence is more dubious, since space plasmas typically have a large number of degrees of freedom. These tools, however, may be informative for the study of low-order deterministic phenomena. Mutual information concepts are useful for correlating different physical parameters and for detecting information flows between different scales. Likewise, soliton transforms are promising tools for analyzing solitary wave packets as those observed by the FREJA and VIKING missions. What is more appealing, however, is the idea of making a quantitative and systematic comparison of the statistical properties (i.e., the dimensionality, the soliton eigenmodes or the bispectrum) as measured in different regions of space.

**Acknowledgments**   It is a pleasure to acknowledge stimulating discussions with V. V. Krasnosel'skikh, F. Lefeuvre, A. W. Wernik, and L. J. C. Woolliscroft. I also gratefully acknowledge the Swiss National Science Foundation for financial support and the Laboratoire de Physique et Chimie de l'Environnement (CNRS, Orléans) where most of this review was written.
    This text is an updated and slightly expanded version of a review that has appeared in : R.E. Stone (ed.) [1996], *The URSI Review on Radio Science 1993-1995*, Oxford, Oxford University Press.